\documentclass[sigconf]{acmart}

\AtBeginDocument{%
  }

\copyrightyear{2025}
\acmYear{2025}
\setcopyright{acmlicensed}\acmConference[MM '25]{Proceedings of the 33rd ACM International Conference on Multimedia}{October 27--31, 2025}{Dublin, Ireland}
\acmBooktitle{Proceedings of the 33rd ACM International Conference on Multimedia (MM '25), October 27--31, 2025, Dublin, Ireland}
\acmDOI{10.1145/3746027.3755513}
\acmISBN{979-8-4007-2035-2/2025/10}

\begin{document}


\title{Tractography-Guided Dual-Label Collaborative Learning for Multi-Modal Cranial Nerves Parcellation}


\author{Lei Xie}
\affiliation{%
  \institution{Zhejiang University of Technology}
  \city{Hangzhou}
  \state{Zhejiang}
  \country{China}}
\email{xielei@zjut.edu.cn}

\author{Junxiong Huang}
\affiliation{%
 \institution{Zhejiang University of Technology}
 \city{Hangzhou}
 \state{Zhejiang}
 \country{China}}
\email{211124030095@zjut.edu.cn}

\author{Yuanjing Feng}
\authornote{Corresponding authors}
\affiliation{%
 \institution{Zhejiang University of Technology}
 \city{Hangzhou}
 \state{Zhejiang}
 \country{China}}
\email{fyjing@zjut.edu.cn}

\author{Qingrun Zeng}
\authornotemark[1]
\affiliation{%
 \institution{Zhejiang University of Technology}
 \city{Hangzhou}
 \state{Zhejiang}
 \country{China}}
\email{superzeng@zjut.edu.cn}

\begin{abstract}
The parcellation of Cranial Nerves (CNs) serves as a crucial quantitative methodology for evaluating the morphological characteristics and anatomical pathways of specific CNs. Multi-modal CNs parcellation networks have achieved promising segmentation performance, which combine structural Magnetic Resonance Imaging (MRI) and diffusion MRI. However, insufficient exploration of diffusion MRI information has led to low performance of existing multi-modal fusion. In this work, we propose a tractography-guided Dual-label Collaborative Learning Network (DCLNet) for multi-modal CNs parcellation. The key contribution of our DCLNet is the introduction of coarse labels of CNs obtained from fiber tractography through CN atlas, and collaborative learning with precise labels annotated by experts. Meanwhile, we introduce a Modality-adaptive Encoder Module (MEM)  to achieve soft information swapping between structural MRI and diffusion MRI. Extensive experiments conducted on the publicly available Human Connectome Project (HCP) dataset demonstrate performance improvements compared to single-label network. This systematic validation underscores the effectiveness of dual-label strategies in addressing inherent ambiguities in CNs parcellation tasks.
\end{abstract}



\maketitle

\section{Introduction}
Cranial nerves (CNs) are essential for sensory functions such as hearing, smell, vision, and taste, and they facilitate non-verbal emotional expression through facial movements \cite{1_yoshino2016visualization}. Due to their delicate nature, careful handling of CNs during neurosurgical procedures is vital to prevent complications that could significantly impact a patient's quality of life \cite{2_hodaie2010vivo,3_jacquesson2019overcoming}. Preoperative parcellation of CNs tracts allows for the visualization of their spatial relationships with adjacent structures like tumors or lesions, which is crucial for accurate diagnosis and effective treatment planning \cite{4_sultana2017mri}\cite{5_xie2023cntseg}.

Traditionally, parcellation of CNs based on diffusion tractography relies on manual curation of streamlines to reconstruct representative pathways \cite{6_zolal2016comparison,7_jacquesson2019probabilistic,8_xie2024anatomy,9_hu2024preoperative}. This process demands significant expertise and time investment, as exemplified by region-of-interest (ROI) selection strategies. In these strategies, trained neuroanatomists interactively identify CN structures by sequentially placing ROIs along desired anatomical trajectories \cite{10_xie2020anatomical,11_he2021comparison}. To overcome operator-dependent variability, automated atlas-based methodologies have emerged as a paradigm shift. These techniques leverage preconstructed neural atlases to algorithmically cluster whole-brain tractograms into anatomically defined CN bundles. This approach eliminates the need for per-subject ROI placement while enhancing consistent anatomical fidelity \cite{12_zeng2023automated,16_decroocq2022automation,11_he2021comparison,15_zeng2021automated,13_huang2022automatic,14_zhang2020creation}.

Recent advancements in voxel-based analysis methods have significantly improved neural parcellation by utilizing various MRI modalities to classify voxels based on associated fiber bundles. For example, Ronneberger et al. \cite{intro_wasserthal2018tractseg} developed TractSeg, a convolutional neural network-based approach that performs fast and accurate volumetric white matter tract parcellation directly from fiber orientation distribution Peaks images, eliminating the need for traditional tractography. Avital et al. \cite{intro_19_avital2019neural} introduced a multimodal fusion framework (AGYnet) for neural segmentation by integrating T1w and Directionally Encoded Color (DEC) images using a Y-net network architecture, effectively combining structural and diffusion information. Similarly, MMFnet \cite{intro_20_xie2023deep} was designed for visual neural pathway parcellation, leveraging T1w and Fractional Anisotropy (FA) images within a multimodal fusion network to enhance parcellation performance. Further, Diakite et al. \cite{diakite2025dual} proposed a modality-relevant feature extraction network to extract T1w and FA images information for optic nerve segmentation. Building upon these developments, CNTSeg \cite{5_xie2023cntseg} has pioneered the integration of multimodal fusion in CNs parcellation by combining structural MRI (T1w images) and diffusion MRI (FA and Peaks images).

However, these voxel-based analysis methods rely heavily on experts on images and the gold standard for manual annotation. Due to the small size of the CNs, the accuracy of manual annotation may not be reliable, and voxel-based methods that rely solely on manually annotated labels are also prone to enter the bottleneck. In addition, due to the differing generative characteristics of various medical imaging modalities, the information they contain is also inconsistent. If training is conducted using only a subset of modalities, the results may be suboptimal due to missing information. In contrast, the neural atlas is generated with reference to all modalities and contains rich structural and directional information, which can supplement the input modalities of the network. Therefore, we want to introduce a new kind of learning method that uses the gold standard as the main reference and incorporates labels generated by automatically annotated neural atlas to improve parcellation performance.

In this paper, we propose DCLNet, a tractography-guided dual-label collaborative learning network for multi-modal cranial nerves parcellation. The most important key point of the network is the introduction of coarse labels generated by the neural atlas. Specifically, we matched the tractography results of the CNs with the neural atlas to obtain three-dimensional streamline data of five pairs of major CNs and then mapped the streamline data to labels on MRI images through registration. Then the DCLNet starts learning by using both types of labels simultaneously. During the learning process, we use T1w images and FA images as multi-modal information and quickly complete the parcellation of five pairs of major CNs.

The main contributions are summarized as follows:
\begin{itemize}
\item We propose a novel multi-modal CNs parcellation framework with tractography-guided dual-label collaborative learning strategy.
\item We introduce a modality-adaptive encoder module (MEM) to achieve soft information swapping between structural MRI and diffusion MRI. 
\item The experimental results demonstrate superiority on HCP dataset compared with other state-of-the-art CNs parcellation methods.
\end{itemize}

\begin{figure*}[ht]
    \centering
    \includegraphics[width=1\linewidth]{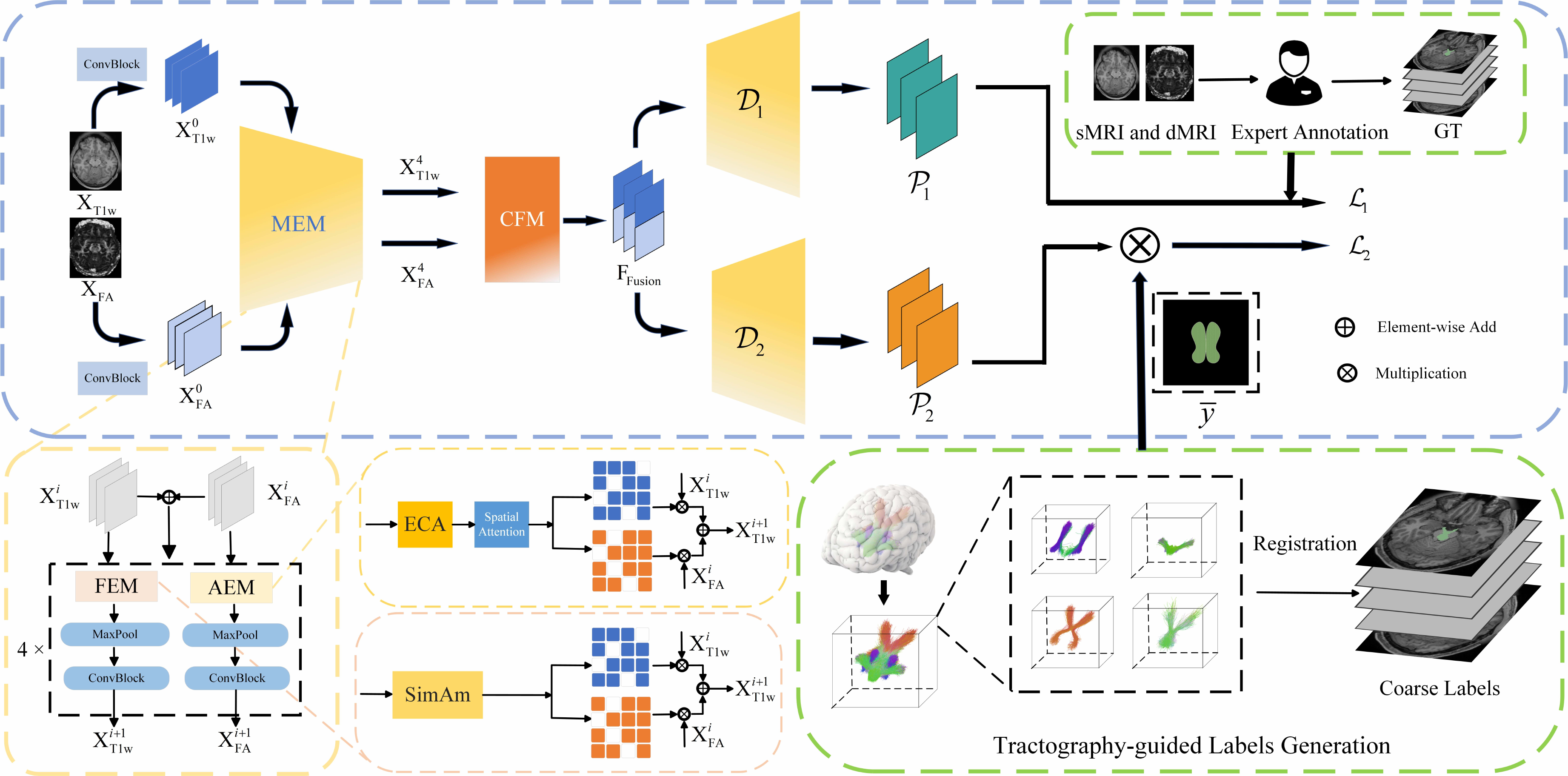}
    \caption{Overview of DCLNet. The top diagram illustrates the overall architecture of DCLNet, which takes T1w and FA images as inputs. The modality-adaptive encoder module (MEM) and cross-fusion module (CFM) modules are employed to fuse information from the two modalities, followed by two decoders corresponding to the dual-label processing pipelines. The MEM module and the process of generating coarse labels via the neural atlas are shown in the four sub-diagrams at the bottom.}
    \Description{DCLNet consists of three major modules: Modality-Adaptive Encoder Moudle (MEM), Cross-Fusion Moudle, and Dual-label Collaborative Learning Moudle.}
    \label{fig:overview}
\end{figure*}

\section{Related Work}
\subsection{Multi-Modal CNs Parcellation}
Early methods for CNs tract parcellation primarily employed deformable models for feature extraction \cite{4_sultana2017mri,17_sultana2016towards,21_sultana2019medial,22_mansoor2015partitioned}. With advancements in deep learning, there has been a notable shift towards deep learning-based models, yielding remarkable results.
For example,  Dolz et al. \cite{23_dolz2017deep} proposed a deep learning classification scheme that uses enhanced features to segment organs at risk in the CN II region in patients with brain cancer. Another approach introduced a CN II parcellation mechanism guided by deep learning features to effectively utilize multimodal structural MRI information \cite{24_mansoor2016deep}. Futher, Xie et al. \cite{5_xie2023cntseg} introduced CNTSeg, a multimodal fusion network for CNs parcellation that combines structural MRI and diffusion MRI data. This network achieved the first parcellation of five pairs of major CNs, including the complex structures of CN III and CN V. Despite the progress made in CNTSeg, how to effectively explore information from structural and diffusion MR has been the focus of research on neural segmentation.

\subsection{CNs Tractography}
Early studies employed ROIs selection strategies to manually extract anatomically relevant tracts from streamlines generated by fiber tractography algorithms~\cite{dyrby2007validation,malcolm2010filtered,aydogan2020parallel}. 
A critical limitation of ROI-based CN identification pipelines stems from their dual reliance on expertise in dMRI tractography and neuroanatomical knowledge. To address these issues, CN atlases were created by fiber clustering to automatically map the pathways, such as the CN II atlas~\cite{12_zeng2023automated}, CN III atlas~\cite{13_huang2022automatic}, CN V atlas~\cite{14_zhang2020creation}, CN VII/VIII atlas~\cite{15_zeng2021automated}, which obtained fiber clustering maps by analyzing the spatial distribution and distance features of different fiber bundles. Diffusion MRI tractography has been successfully applied to CN identification, offering the advantage of non-invasive in vivo mapping of three-dimensional trajectories~\cite{jacquesson2019overcoming,jacquesson2019full,hu2024preoperative,xie2024anatomy}. Therefore, incorporating the 3D trajectories of CNs derived from diffusion tractography into a voxel-based multi-modal framework is key to enhancing the accuracy of CNs parcellation. In this paper, we employ neural mapping as a priori to design collaborative learning algorithms that co-optimize the model using both gold-standard annotations and coarse labels.


\section{Methodology}
In this section, we first introduce the overview of the DCLNet. Then, we present the main modules of the feature extraction process. Finally, we describe the dual-label training strategies.

\subsection{Dual-label Collaborative Learning Network}
The network architecture of the proposed DCLNet is shown in  \autoref{fig:overview}, comprising a modality-adaptive encoder module (MEM), a cross-fusion module (CFM) and a dual-label collaborative learning module. The modality-adaptive encoder module is designed to facilitate preliminary information exchange between input modalities, then output features enter the cross-fusion module for further information exchange. The cross-fusion module utilizes the cross-attention mechanism to exchange information between features and output concatenated features. The concatenated features will enter two decoders, which connect the processing pipelines of two different labels. Both decoders are U-shaped decoders, it is worth noting that the second decoder incorporates a dropout layer to prevent overfitting. 

As shown in \autoref{fig:overview}, the input T1w modality $\mathrm{X}_{\text{T1w}}\in \mathbb{R}^{1 \times H \times W}$ and FA modality $\mathrm{X}_{\text{FA}}\in \mathbb{R}^{1 \times H \times W}$ will get  T1w feature $\mathrm{X}^0_{\text{T1w}}$ and FA feature $\mathrm{X}^0_{\text{FA}}$ after the convolutional layer, and they will be used as inputs to the MEM for feature exchange fusion, and the final output features $\mathrm{X}^4_{\text{T1w}}$ and $\mathrm{X}^4_{\text{FA}}$, as:
\begin{equation}
\mathrm{X}^4_{\text{T1w}}, \mathrm{X}^4_{\text{FA}} = \mathrm{MEM}(\mathrm{X}^0_{\text{T1w}}, \mathrm{X}^0_{\text{FA}})
\end{equation}
where $\mathrm{MEM}(\cdot)$ means the introduced MEM module. Then, $\mathrm{X}^4_{\text{T1w}}$ and $\mathrm{X}^4_{\text{FA}}$ will be sent to the cross-fusion module to get fused features $\mathrm{F}_{\text{Fusion}}$, as:
\begin{equation}
\mathrm{F}_{\text{Fusion}} = \mathrm{CFM}(\mathrm{X}^4_{\text{T1w}}, \mathrm{X}^4_{\text{FA}})
\end{equation}
where $\mathrm{CFM}(\cdot)$ means the used cross-fusion module. Then, the fused features $\mathrm{F}_{\text{Fusion}}$ are restored by two decoders ($\mathcal{D}_1$ and $\mathcal{D}_2$), which have the same structure except that one of them has a random dropout layer added. These operations can be formulated as follows: 
\begin{equation}
\mathcal{P}_{1} = \mathrm{Sigmoid}(\mathrm{Conv}_{1 \times 1}(\mathcal{D}_1(\mathrm{F}_{\text{Fusion}})))
\end{equation}
\begin{equation}
\mathcal{P}_{2} = \mathrm{Sigmoid}(\mathrm{Conv}_{1 \times 1}(\mathcal{D}_2(\mathrm{F}_{\text{Fusion}})))
\end{equation}
where the $\mathrm{Conv}{_{1 \times 1}}(\cdot)$ means a convolution layer with a kernel size of $1 \times 1$, $\mathrm{Sigmoid}(\cdot)$ denotes the Sigmoid activation function.
After obtaining $\mathcal{P}_{1}$ and $\mathcal{P}_{2}$, we use dual-label collaborative training strategy to train the model, which consists of both the expert-annotated ground truth processing pipeline and the coarse labels processing pipeline processing flows. The total loss function of the our DCLNet is defined as:
\begin{equation}
\mathcal{L}_{total} = \mathcal{L}_{1}(\mathcal{P}_{1},\mathcal{G}) + \mathcal{L}_{2}(\bar y\otimes\mathcal{P}_{2},\mathcal{G})
\end{equation}
where $\mathcal{G}$ are ground truth with expert manual labeling \cite{5_xie2023cntseg}, and $\otimes$ means the element-wise multiplication. $\bar {y}$ represents the coarse labels generated from the atlas guided by fiber tractography.  Since the coarse labels incorporate referential information from the atlas, the samples within $\bar {y}$ are of high-confidence prediction results. We compute the loss for these pixels by comparing them with the gold standard and incorporate this into the overall loss of the network. 
For the expert-annotated ground truth processing pipeline, we use a combined loss function $\mathcal{L}_{1}$ of Dice loss ($\mathcal{L}_{\text{Dice}}$) and BCE loss ($\mathcal{L}_{\text{BCE}}$):
\begin{equation}
\mathcal{L}_{1} = \mathcal{L}_{\text{Dice}} + \mathcal{L}_{\text{BCE}}
\end{equation}
where the Dice loss is defined as:
\begin{equation}
\mathcal{L}_{\text{Dice}} = 1 - \frac{2 \sum y \cdot \hat{y}}{\sum y + \sum \hat{y}}
\end{equation}
where \(\hat{y}\) represents the predicted values, and \(y\) represents the ground truth labels.
The BCE (Binary Cross-Entropy) loss is defined as:
\begin{equation}
\mathcal{L}_{\text{BCE}} = -\sum \left( y \log \hat{y} + (1 - y) \log (1 - \hat{y}) \right)
\end{equation}
For the coarse-labels processing pipeline, we use the coarse labels generated from the atlas to generate a binary mask, where pixel values inside the mask are set to 1 and those outside are set to 0. We use the BCE loss $\mathcal{L}_{\text{BCE}}$ as the loss function $\mathcal{L}_{2}$ for the coarse-labels processing pipeline.

\subsection{Modality-Adaptive Encoder Module}
In this paper, we introduce the modality-adaptive encoder module (MEM) \cite{han2024modality} to enhance the feature extraction process while minimizing the amount of introduced parameters, which contains the fixed exchange module (FEM) and the adaptive exchange module (AEM). The specific details of MEM are given as follows.

At the $i^{th}|i \in \{0, 1, 2, 3\}$ step, given the T1w feature $\mathrm{X}^i_{\text{T1w}}$ and the FA feature $\mathrm{X}^i_{\text{FA}}$, the FEM and AEM are applied to facilitate preliminary information exchange. Specifically, in the T1w feature extraction branch, we add the T1w feature $\mathrm{X}^i_{\text{T1w}}$ and FA feature $\mathrm{X}^i_{\text{FA}}$ to obtain the input fused feature $\mathrm{X}^i_{\text{input}}$ of the FEM and AEM, defined as:
\begin{equation}
\mathrm{X}^i_{\text{input}} = \mathrm{X}^i_{\text{T1w}} + \mathrm{X}^i_{\text{FA}}
\end{equation}
where $i \in \{0, 1, 2, 3\}$ indicates that the operation is repeated four times, with $i$ serving as the index for each iteration. We then employ SimAM \cite{simAm_wang2020eca} to compute channel-wise attention coefficients $\mathrm{S}_i$ based on the fused input feature $\mathrm{X}^i_{\text{input}}$, formulated as:
\begin{equation}
\mathrm{S}_i = \mathrm{SimAM}(\mathrm{X}^i_{\text{input}})
\end{equation}
where $\mathrm{SimAM}(\cdot)$ means the operation of applying lightweight attention mechanism SimAM to input feature maps in FEM module. And the output $\mathrm{X}^{i+1}_{\text{T1w}}$ of FEM for T1w modality becomes:
\begin{equation}
\mathrm{X}^{i+1}_{\text{T1w}} = \mathrm{Conv}(\mathrm{MP}(\mathrm{S}_i\mathrm{X}^i_{\text{T1w}} +(1-\mathrm{S}_i) \mathrm{X}^i_{\text{FA}})) 
\end{equation}
where $\mathrm{MP}(\cdot)$ denotes the max-pooling operations, and $\mathrm{Conv}(\cdot)$ represents convolutional block operation, which includes two $3 \times 3$ convolutional layers and two batch normalization layers. For FA feature extraction branch, we introduce another lightweight and learnable exchange module, AEM, which integrates efficient channel attention (ECA) \cite{wang2020eca} with spatial attention (SA) mechanisms. The attention-enhanced coefficient map $\mathrm{E}_i$ is computed as:
\begin{equation}
\mathrm{E}_i = \mathrm{SA}(\mathrm{ECA}(\mathrm{X}^i_{\text{input}}))
\end{equation}
where $\mathrm{ECA}(\cdot)$ and $\mathrm{SA}(\cdot)$ stand for the efficient channel attention and spatial attention mechanisms. And we obtain the output $\mathrm{X}^{i+1}_{\text{FA}}$ of the FA modality:
\begin{equation}
\mathrm{X}^{i+1}_{\text{FA}} = \mathrm{Conv}(\mathrm{MP}(\mathrm{E}_i\mathrm{X}^i_{\text{T1w}} +(1-\mathrm{E}_i) \mathrm{X}^i_{\text{FA}})) 
\end{equation}
The feature maps, which now contain fused information from both modalities, are further processed with convolution operations, and this process is repeated four times. Finally, we obtain the modality-specific features $\mathrm{X}^4_{\text{T1w}}$ and $\mathrm{X}^4_{\text{FA}}$.

\subsection{Tractography-guided Labels Generation}
The labels generated based on diffusion tractography atlas can represent rough positional information of CNs.  In previous work\cite{11_he2021comparison,15_zeng2021automated,13_huang2022automatic,14_zhang2020creation,xie2025automatedmappingpathwayscranial}, individual neural atlas for CN II, CN III, CN V, and CN VII/VIII have been created separately. Inspired by this, we take these neural atlas as  priori knowledge and conduct research. The generation of these labels mainly involves three steps: neural atlas creation, voxel-based region generation, and subject-specific registration. 

 Neural atlas creation: First, we use tractography data from 50 subjects to align into a common space using entropy-based groupwise registration, combining affine and multi-scale B-spline transformations. Then proceed with a two-stage clustering process. In stage 1, spectral clustering divides fibers into 6,000 clusters. Outliers are filtered, and 106 CNs-related clusters are identified via ROI-based screening. In stage 2, enhanced clustering merges and refines these clusters into 200 subgroups. Expert annotation selects 74 anatomically validated clusters, representing 5 CNs pairs. Each streamline from the subject-specific CNs tractography is registered and assigned to the nearest cluster in the multi-stage fiber atlas. Outlier fibers are filtered out using the same parameters applied during the creation of the atlas. Finally, automated CNs identification for the new subject is achieved by matching the subject-specific clusters to the corresponding clusters defined in the atlas.
 
Voxel-based region generation: The constructed neural atlas exists in the form of three-dimensional streamlines, so it is converted into a voxel-based CN region. First, we use the \textit{tckmap} command in the \textit{MRtrix3} toolkit \cite{tournier2012mrtrix} to map high-dimensional data to the voxel level of the images. Since CN structures are continuous, we remove some isolated islands that deviate from the main region, ultimately obtaining a voxel-based neural atlas.

Subject-specific registration: To obtain the final labels, we use \textit{FSL} software \cite{jenkinson2012fsl} to register the voxel-based neural atlas to the individual space. Our neural atlas can be applied to any new individual, which allows the proposed DCLNet to be applied to other different datasets.

\section{Experiments}
\subsection{Datasets}
We used an open-source dataset from the Human Connectome Project (HCP). A total of 102 subjects were selected, all scanned at Washington University in St. Louis using a customized Siemens Skyra 3T scanner (Siemens AG, Erlangen, Germany).The diffusion MRI (dMRI) acquisition protocol included 18 baseline images with \( b = 0\ \mathrm{s/mm}^2 \) and 270 diffusion-weighted volumes with three different \( b \)-values: 1000, 2000, and 3000 \( \mathrm{s/mm}^2 \). The scanning parameters were: TR = 5520 ms, TE = 89.5 ms, matrix size = \( 145 \times 174 \times 145 \), and voxel resolution = \( 1.25 \times 1.25 \times 1.25\ \mathrm{mm}^3 \). FA images were computed using the MRtrix3 toolbox.The structural T1w images were acquired with the following parameters: TR = 2400 ms, TE = 2.14 ms, matrix size = \( 145 \times 174 \times 145 \), and resolution = \( 1.25 \times 1.25 \times 1.25\ \mathrm{mm}^3 \). The dataset provides both high-resolution T1w and preprocessed dMRI data.
For each of the 102 subjects, reference annotations for five pairs of cranial nerves were generated by projecting the corresponding fiber tract streamlines onto voxel-based binary segmentation maps.

\subsection{Implementation Details}
All experiments were conducted on two parallel NVIDIA RTX 3060 GPUs. The network was implemented using Python 3.9.7, PyTorch 2.0.1, Torchvision 0.15.2, and CUDA 12.1. A batch size of 32 and an initial learning rate of 0.002 were used. The model was trained for 200 epochs using the SGD optimizer, and the best-performing weights were selected based on validation performance. During training, 2D axial slices of the input volumes were used. Data augmentation techniques included random horizontal flipping and random modifications to brightness, contrast, and hue. The structural MRI data from the HCP dataset were resampled to a resolution of $128 \times 160 \times 128$ without removing any brain tissue. The dMRI data were resized to match the same dimensions. A total of 102 subjects from the HCP dataset were used, and 5-fold cross-validation was performed with a 4:1 split between training and validation sets.

\subsection{Evaluation Metrics}
We assessed the performance of the model segmentation using various metrics, including dice, jaccard, precision, and average hausdorff distance (AHD) \cite{5_xie2023cntseg}.

Dice Coefficient (DICE): Measures the spatial overlap between the predicted parcellation area A and ground truth B.
\begin{equation}
Dice(A, B) = \frac{2|A \cap B|}{|A| + |B|} = \frac{2TP}{2TP + FP + FN}
\end{equation}

Jaccard Index (IoU or Jaccard): Similar to Dice, but emphasizes overlap relative to the union of A and B.
\begin{equation}
Jaccard(A, B) = \frac{|A \cap B|}{|A \cup B|} = \frac{TP}{TP + FP + FN}
\end{equation}

Precision: Measures the accuracy of positive predictions by evaluating the ratio of true positives (TP) to all predicted positives (TP + FP).
\begin{equation}
Precision(A, B) = \frac{|A \cap B|}{|A|}   =\frac{TP}{TP + FP}
\end{equation}

Average Hausdorff Distance (AHD): Quantifies shape similarity by measuring the average Euclidean distance between the surfaces of A and B. Lower values indicate better boundary alignment.
\begin{equation}
\text{AHD}(A, B) = \frac{1}{2} \left( \frac{1}{|A|} \sum_{a \in A} \min_{b \in B} \| a - b \| + \frac{1}{|B|} \sum_{b \in B} \min_{a \in A} \| b - a \| \right)
\end{equation}

\subsection{Comparison with State-of-the-Art Methods}
We compared DCLNet with four state-of-the-art (SOTA) deep learning based parcellation models on the HCP dataset. 
The networks used are TractSeg\cite{intro_wasserthal2018tractseg}, AGYNet\cite{intro_19_avital2019neural}, MMFNet\cite{intro_20_xie2023deep}, and CNTSeg\cite{5_xie2023cntseg}. All models were evaluated under the same experimental setup and data parcellation procedures. \autoref{result1} and \autoref{result2} illustrate the quantitative metrics for different models with average results over 5-fold cross-validation on the testing dataset.

\begin{table}[ht]
\caption{Comparison results of our DCLNet and SOTA cranial nerves parcellation methods on HCP dataset. The best-performing results are highlighted in bold.}
\label{result1}
\resizebox{\columnwidth}{!}{
\begin{tabular}{ccccc}
\toprule
                             & Dice{[}\%{]}↑                     & Jaccard{[}\%{]}↑            & Precision{[}\%{]}↑      & AHD↓          \\
\midrule
TractSeg                     & 65.39±5.23                     & 49.12±5.39          & 66.98±5.22          & 0.461±0.123          \\
AGYnet                       & 70.41±4.34                     & 55.33±4.57          & 71.02±4.41          & 0.390±0.130          \\
MMFNet                       & 71.94±3.14                     & 57.03±3.62          & 72.43±4.49          & 0.372±0.084          \\
CNTSeg                       & 71.59±4.24                     & 55.88±4.89          & 72.07±4.17          & 0.381±0.113          \\
DCLNet                     & \textbf{72.60±3.45}            & \textbf{57.90±3.99} & \textbf{73.44±4.51}   & \textbf{0.357±0.079} \\
\bottomrule
\end{tabular}
}
\end{table}

\begin{table*}[ht]
\footnotesize
\caption{Comparison of 5 pairs of CNs parcellation performance}
\label{result2}
\resizebox{\textwidth}{!}{
\begin{tabular}{cccccc|cccccc}
\hline
\multicolumn{1}{l}{} & \multicolumn{5}{c|}{\textbf{Dice{[}\%{]}↑}}                                                                  & \multicolumn{1}{l}{} & \multicolumn{5}{c}{\textbf{Jaccard{[}\%{]}↑}}                                                                       \\ \hline
\multicolumn{1}{l}{} & CN II               & CN III              & CN V                & CN VII/VIII          & Mean                &                      & CN II                & CN III               & CN V                 & CN VII/VIII          & Mean                 \\ \hline
TractSeg              & 72.18±5.58          & 65.26±6.36          & 61.09±7.97          & 63.02±8.71           & 65.39±5.23          & TractSeg             & 56.75±6.51           & 48.76±6.88           & 44.42±7.78           & 46.56±8.72           & 49.12±5.39           \\
AGYnet                & 84.23±2.76          & 63.73±6.56          & 63.63±6.38          & 70.03±8.44           & 70.41±4.34          & AGYnet               & 72.85±4.01           & 47.08±6.78           & 46.98±6.83           & 54.42±8.47           & 55.33±4.57           \\
MMFNet                & 84.41±3.16          & 66.01±5.88          & 64.84±5.35          & \textbf{72.51±6.04}  & 71.94±3.14          & MMFNet               & 73.15±4.53           & 49.55±6.47           & 48.20±5.71           & \textbf{57.22±7.31}  & 57.03±3.62           \\
CNTSeg                & 84.35±2.97          & 66.74±6.47          & 63.69±7.49          & 68.78±8.04           & 71.59±4.24          & CNTSeg               & 73.04±4.26           & 50.42±7.03           & 47.14±7.66           & 52.94±8.65           & 55.88±4.89           \\
DCLNet                & \textbf{85.72±2.89} & \textbf{67.16±6.42} & \textbf{65.01±5.58} & 72.49±5.94           & \textbf{72.60±3.45} & DCLNet               & \textbf{75.12±4.24}  & \textbf{50.89±6.90}  & \textbf{48.41±6.09}  & 57.18±7.17           & \textbf{57.90±3.99}  \\ \hline
\multicolumn{1}{l}{} & \multicolumn{5}{c|}{\textbf{Precision{[}\%{]}↑}}                                                             & \multicolumn{1}{l}{} & \multicolumn{5}{c}{\textbf{AHD↓}}                                                                               \\ \hline
\multicolumn{1}{l}{} & CN II               & CN III              & CN V                & CN VII/VIII          & Mean                &                      & CN II                & CN III               & CN V                 & CN VII/VIII          & Mean                 \\ \hline
TractSeg              & 73.29±6.47          & 66.38±9.06          & 61.57±10.68         & 66.69±11.60           & 66.98±5.22          & TractSeg             & 0.365±0.197          & 0.442±0.156          & 0.568±0.188          & 0.469±0.218          & 0.461±0.123          \\
AGYnet                & 83.97±5.32          & 63.29±8.73          & 63.58±9.82          & 73.22±11.98          & 71.02±4.41          & AGYnet               & 0.175±0.053          & 0.476±0.175          & 0.544±0.226          & 0.363±0.237          & 0.390±0.130           \\
MMFNet                & 84.02±5.61          & \textbf{68.82±8.19} & 63.61±9.87          & 73.26±10.99          & 72.43±4.49          & MMFNet               & 0.176±0.069          & 0.471±0.195          & \textbf{0.514±0.163} & 0.328±0.126          & 0.372±0.084          \\
CNTSeg                & 85.56±5.22          & 68.32±8.72          & \textbf{65.22±9.96} & 69.18±11.08          & 72.07±4.17          & CNTSeg               & 0.170±0.050          & 0.445±0.257          & 0.540±0.217          & 0.370±0.152          & 0.381±0.113          \\
DCLNet                & \textbf{86.16±5.48} & 68.16±8.94          & 64.60±9.05          & \textbf{74.83±11.00} & \textbf{73.44±4.51} & DCLNet               & \textbf{0.166±0.070} & \textbf{0.419±0.160} & 0.522±0.159          & \textbf{0.322±0.119} & \textbf{0.357±0.079} \\ \hline
\end{tabular}}
\end{table*}

\begin{figure*}[htbp]
    \centering
    \includegraphics[width=1\linewidth]{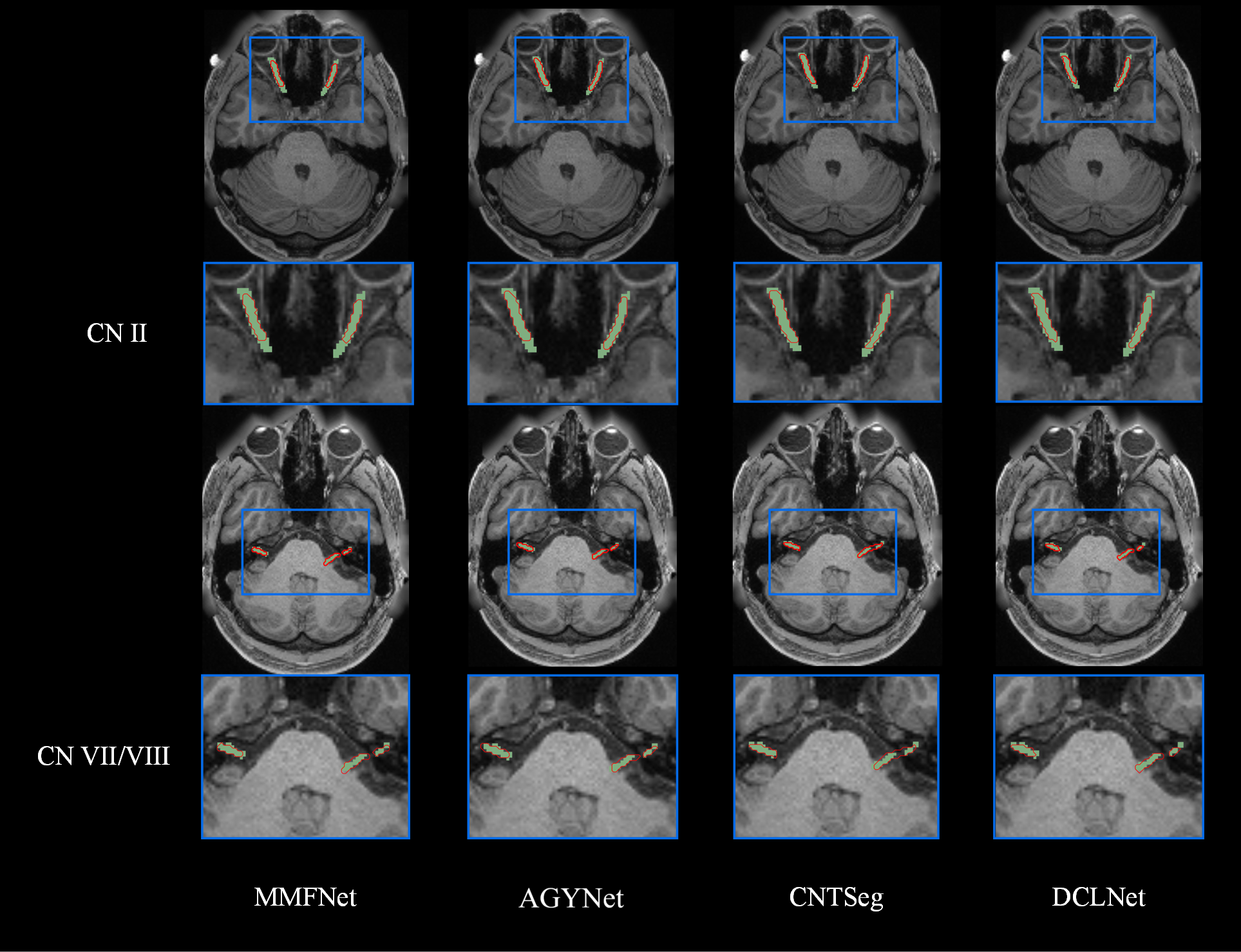}

    \caption{Qualitative comparison of results on HCP No. 100206 subject, the green areas in the figure indicate the results labeled by the experts, and the red indicates the areas predicted by each method.}
    \Description{This figure shows qualitative results of DCLNet.}
    \label{visual}
\end{figure*}

\begin{table}
\caption{Results of ablation experiments. $\checkmark$ indicates the presence of this block, with the best results highlighted. $\checkmark^{1}$ and $\checkmark^{2}$ stand for MEM without AEM and without FEM, respectively.}
\label{ablation results}
\resizebox{\columnwidth}{!}{
\begin{tabular}{ccc|cccc}
\hline
\textbf{Baseline} & \textbf{MEM} & \textbf{DCL} & \textbf{Dice {[}\%{]}↑} & \textbf{Jaccard {[}\%{]}↑} & \textbf{Precision{[}\%{]}↑} & \textbf{AHD↓}        \\ \hline
\checkmark                 &              &                            & 71.94±3.14             & 57.03±3.62                & 72.43±4.49                  & 0.372±0.084          \\
\checkmark                & $\checkmark^{1}$          &     &72.16±3.62
                       &      57.39±4.14
       &      \textbf{73.55±4.46}
          &   0.359±0.080
                 \\
\checkmark                & $\checkmark^{2}$      &                            &   72.46±3.28
           &   57.70±3.81
            &  73.38±4.25
        &   0.359±0.076
      \\
\checkmark                & \checkmark           &                            & 72.33±3.56             & 57.58±4.10                & 73.47±4.56         & 0.361±0.082          \\
\checkmark                & \checkmark           & \checkmark                         & \textbf{72.60±3.45}    & \textbf{57.90±3.99}       & 73.44±4.51                  & \textbf{0.357±0.079} \\ \hline
\end{tabular}
}
\end{table}
Except for CNTSeg, the other comparison methods are not specifically designed for cranial nerves (CNs) parcellation. To ensure a fair and meaningful comparison, we implemented the parcellation of CNs in accordance with the specific modalities each method utilizes and the network architectures they employ. 
Specifically, TractSeg focuses on white matter tract parcellation, utilizing Peaks images fed into a CNN network. 
AGYnet specializes in nerve parcellation, processing T1w images and DEC images through a Y-net network. 
Meanwhile, MMFnet is designed for the visual neural pathway
parcellation, leveraging T1w images and FA images in a
multimodal fusion network. 
All models, including the baseline network and our DCLNet, were trained under identical hardware settings and dataset splits. 
\autoref{result1} reports the parcellation results of our DCLNet and
the competing methods for each CNs tract in terms of Dice, 
Jaccard, Precision, and AHD metrics. 
The best-performing scores are highlighted for each metric. As demonstrated
in \autoref{result1}, DCLNet outperforms all the CNs tract
parcellation models and achieves state-of-the-art (SOTA) performance. For
instance, our DCLNet segments five pairs of CNs with the
mean Dice, mean Jaccard, mean Precision, and mean AHD of 
72.60\%, 57.90\%, 73.44\% and 0.357, which are higher than baseline MMFnet by 0.66\%, 0.87\%, 
1.01\%, and 0.015. Compared with CNTSeg, which is specifically designed for cranial nerves (CNs) parcellation, our DCLNet achieves improvements of 1.01\%, 2.02\%, 1.37\%, and 0.024, respectively, thereby demonstrating the effectiveness of our method for CN parcellation tasks.

To clearly demonstrate the performance of our model, we have listed five specific pairs of CNs parcellation results in \autoref{result2}. As shown in \autoref{result2}, our method significantly outperforms other networks in the parcellation task of CN II, achieving SOTA performance. Moreover, DCLNet also demonstrates leading results in the parcellation of CN III, CN V, and CN VII/VIII. These results validate the effectiveness and advantages of DCLNet in the parcellation of individual cranial nerves.

To provide a more intuitive illustration of the performance of our method, we select subject No. 100206 from the HCP dataset to represent the CNs parcellation results. \autoref{visual} shows the qualitative CN II and CN VII/VIII parcellation results generated by different methods. The green areas in the figure represent expert annotations, while the red areas correspond to the predictions of each method. As shown in \autoref{visual}, our method demonstrates greater overlap with the ground truth compared to other SOTA methods. Notably, it achieves superior accuracy in detecting regions near anatomical boundaries and in cases where the labels are non-contiguous. For instance, the CN VII/VIII prediction results from the AGYNet and CNTSeg show that, under discontinuous labeling conditions, a large number of incorrect samples are falsely identified as positive. This highlights the robustness of our method in handling fine anatomical structures.

\subsection{Ablation Study}
\subsubsection{Effectiveness on Components of DCLNet}
The results of the ablation experiments set up in this paper
are shown in \autoref{ablation results}. Compared to the baseline network MMFNet, under a single-label learning scenario, individually adding either the AEM or FEM module, or using  the MEM module that combines both, significantly improves performance metrics. Specifically, when MEM was added, performance increased by 0.39\%, 0.55\%, 1.04\%, 0.011 in Dice, Jaccard, Precision, and AHD metrics. Meanwhile, the introduction of both MEM and DCL modules increased these indicators by 0.66\%, 0.87\%, 1.01\%, and 0.015. It is noteworthy that although Precision shows improvement, the overall performance decreases when training with dual labels. This occurs because precision primarily measures the proportion of true positive predictions among all positive predictions. When the prediction map has smaller coverage and is fully contained within the ground truth area, it artificially inflates the number of true positives (by minimizing false positives) but may severely increase false negatives. This trade-off leads to higher Precision.

\subsubsection{vs. CNTSeg}
CNTSeg, proposed by Xie et al.\cite{5_xie2023cntseg}, represents an initial exploration of cranial nerve tract parcellation using fully convolutional neural networks, incorporating T1w, FA, and Peaks images as inputs. The incorporation of Peaks modality enhanced the input data diversity of CNTSeg, allowing it to achieve SOTA performance at the time. However, the acquisition and processing of Peaks data require additional time and are not commonly prioritized in clinical practice. In contrast, DCLNet relies only on T1w and FA images, resulting in a more lightweight and clinically practical architecture. Quantitatively, DCLNet outperforms CNTSeg in delineation accuracy across all five pairs of cranial nerves. Qualitatively, CNTSeg introduces substantial false positive artifacts when processing intricate or fragmented nerve regions within imaging slices, while DCLNet precisely captures spatial discontinuities and boundary details in targeted areas.

\section{Discussion}
The judgment of cranial nerve region is an important part of the clinical diagnosis of nervous system. An automatic and efficient cranial nerve parcellation method has played a role in improving the accuracy and efficiency of diagnosis. In this work, the proposed DCLNet introduces a more effective information fusion module and the labels obtained by neural atlas to traning process. Experiments on HCP datasets show that our method achieves better performance than the existing methods.

At present, fiber clustering methods are commonly employed in white matter mapping and visualization. These techniques group tractography streamlines based on their geometric trajectories. Unlike traditional approaches that rely on manually defined regions of interest (ROIs), fiber clustering can automatically identify fiber bundles associated with WM or CNs through data-driven model training. Based on the wide applicability of the atlas, we introduced the neural atlas obtained by fiber clustering for learning, which expanded the specific application of fiber bundle tractography. In future work, we will consider how to more fully utilize the results of fiber tracking for learning tasks based on multi-modal information.

However, our method also has some limitations. Firstly, regarding the use of labels, in our experiments they were applied as masks, without fully exploiting their potential or fostering more direct interaction with the gold standard. Secondly, the proposed method was evaluated on the HCP dataset, this database is characterized by high image quality. This poses challenges for direct translation to clinical scenarios, where image quality is typically lower. Consequently, model performance in such settings may require additional fine-tuning. Developing a more generalized and robust model capable of adapting to lower-quality clinical data remains a critical direction for future research.

\section{Conclusion}
In this work, we propose a dual-label collaborative learning network for multi-modal parcellation, specifically designed for delineating five pairs of cranial nerves. The key contribution of the network is the introduction of coarse labels of cranial nerves obtained from tractography through neural atlas, and collaborative learning between coarse labels and expert-annotated precise labels. Extensive experiments conducted on the publicly available HCP dataset demonstrate performance improvements compared to single-label network. This systematic validation underscores the effectiveness of dual-label strategies in addressing inherent ambiguities in cranial nerve parcellation tasks.

\begin{acks}
This work was supported in part by the National Natural Science Foundation of China (No. U22A2040, U23A20334, 62403428); Natural Science Foundation of Zhejiang Province (No. LQ23F030017, LMS25F030004); Zhejiang Province Science and Technology Innovation Leading Talent Program (No. 2021R52004).
\end{acks}

\bibliographystyle{ACM-Reference-Format}
\balance
\bibliography{sample-base}

\appendix

\end{document}